# Enhancing Operational Grid Resilience Against Wildfires Under Decision-Dependent Uncertainties

Arastoo H Salimi, *Student Members, IEEE*, Hamidreza Nazaripouya, *Senior Member, IEEE*

***Abstract*-- This paper proposes a new automated decision-making framework to enhance the resilience of electrical systems against wildfires by applying operational strategies that account for decision-dependent uncertainty (DDU). The proposed framework incorporates both preventive and corrective measures, enabling adaptive and automated decision-making throughout the course of evolving wildfire scenarios. First, a baseline multistage optimization model is presented as a foundation to support wildfire-driven operational decision-making. The model then incorporates DDU, wherein Public Safety Power Shutoff (PSPS) decisions made in earlier stages influence the probabilities and parameters of future wildfire scenarios. To efficiently solve the resulting complex optimization problem, a mathematical decomposition algorithm is employed. The effectiveness of the proposed approach is demonstrated through case studies on the IEEE 30-bus system. Simulation results confirm that incorporating the impact of DDU into the optimization process provides more realistic, and operationally resilient solutions in the face of evolving wildfire threats.**

***Indices and Sets***

| | |
|---|---|
| $e \in E$ | Set of lines. |
| $\beta_i^e$ | Denotes the subset of lines that connected to bus *i*. |
| $g \in G$ | Set of generators. |
| $i \in \beta$, $j \in \beta$ | Set of buses. |
| *a(n)* | Denotes the parent node of *n*. |
| *c(n)* | Denotes the set of child nodes of *n*. |
| *D, G, L,* and $\beta$ | Represent the sets of load demand, generators, transmission lines, and buses, respectively. |
| $\beta_i^D$, $\beta_i^G$, $\beta_i^e$ | Represent the set of loads, generators, and lines connected to bus *i*, respectively. |
| $\psi^{d,n}$ | Weight to express that certain load, or service may be prioritized over others, in node *n*. |
| $\alpha$ | $\alpha \in [0 \;\; 1]$ represents the trade-off between prioritizing load service and cost reduction (low $\alpha$) and avoiding wildfire risk (high $\alpha$). |

***Parameters and constants***

| | |
|---|---|
| $b^{e,n}$ | Susceptance of branch e (line) in the network at node *n*. |
| $D^{i,n}$ | Real power demand at bus *i* (MW), at node *n*. |
| $P_{max}^g$, $P_{min}^g$ | Maximum and minimum real power generation limit at generator $g$ (MW). |
| $R^{fire,n}$ | Denotes the risk of wildfire in node *n*. |
| $D^{Tot,n}$ | Represents the total amount of load delivered in node *n*. |
| $R^{d,n}$ | The wildfire risk associated with load *d* at node *n* in a specific area. |
| $R^{g,n}$ | The wildfire risk associated with generator *g* at node *n* in a specific area. |
| $R^{e,n}$ | The wildfire risk associated with line *e* at node *n* in a specific area. |
| $R^{i,n}$ | The wildfire risk associated with bus *i* at node *n* in a specific area. |
| $C^{g,n}$ | Function to calculate the associated cost for each generator at node *n*. |
| $\theta^{max}$, $\theta^{min}$ | Additional constants added to the angle differences. |

***Notation for Optimization Problem***

| | |
|---|---|
| $Z^1$ | Influential variable in the first stage; if $Z^1$>0, affects scenario weights and objective coefficients. |
| τ | Latency parameter (1 ≤ τ ≤ T−1); constraints of nodes in stages t ≤ τ cannot be changed by influential variables. |
| $Z^n$ | Influential variable at node n ($t_n \le T-\tau$); if $Z^n > 0$, affects constraint coefficients of successors in stage t+τ. |
| $\sigma_\tau^n$ | τ-th ancestor of node n ($t_n > \tau$); denoted as $\sigma$. |
| $I^n$ | Index set for ordered disjoint value intervals of $Z^n$ (includes 0 to represent $Z^n$ not activated). |
| $Z_\iota^n$ | ι-th value interval of $Z^n$ and its upper bound; $\bar{Z}_0^n = 0, \bar{Z}_{-1}^n = 0.001$. |
| $\delta_\iota^n$ | Binary variable: 1 if $Z_\iota^n$ is activated; ensures only one $\delta_\iota^n$ per set $I^n$. |
| $w_\iota^\omega$ | Weight of scenario ω if $Z_\iota^1$ is activated (ω ∈ Ω); sum to 1. |
| $X_{T_1}$ | Binary implementable decision vector at the first stage (aggregation node $T_1$). |
| $Y_{T_1}$ | Continuous implementable decision vector at the first stage (aggregation node $T_1$). |
| $w_0^\omega$ | Original scenario weight if $Z^1$ is not activated. |
| $\rho^X$ | Vector of scaling coefficients for binary variables in the pseudo-gradient update. |
| $\rho^Y$ | Vector of scaling coefficients for continuous variables, obtained from the expected $l_1$-norm deviation relative to the implementable solution. |
| $\mu_s^*$ | Pseudo-gradient vector variables in scenario cluster *s*. |

## I. Introduction

In recent years electrical infrastructures have experienced significant damage and disruption due to large fires. In rural areas, wildfire threatens portions of the transmission grid and can, in some cases, threaten the stability of the grid itself. The 2017 Thomas Fire, for example, caused outages on the Santa

Barbara path, leading to widespread customer service disruptions by intermittently interrupting transmission lines and causing outages for more than 85,000 customers. A report released by the Department of Energy (DOE) summarizes the bulk of power emergencies and disturbances, by their cause as well as number of customers affected. In this study, wildfire, with 933,547 affected customers on average, is at the top of the list [1]. According to the study performed by Lawrence Berkeley National Lab (LBNL), the number and severity of future wildfires is expected to increase rapidly, exposing significant populations and infrastructure to wildfire losses and service disruptions [2]. On the other hand, electrical infrastructure itself plays a key role in the creation and spread of wildfires. Power lines and electrical equipment problems have regularly ranked among the top identified causes of wildfires. California's second-largest wildfire ignited when power lines came into contact with a tree. This fire, which began on July 13, 2021, burned 963,309 acres across five counties in Northern California before being contained. Notably, wildfires sparked by power lines and electrical equipment are among the most destructive wildfires. According to California fire officials, the deadliest wildfire in California's history was caused, at least in part, by transmission lines. This fire burned a total of 153,336 acres, destroying 18,804 structures and resulting in 85 civilian fatalities and several firefighter injuries [3].

Although significant efforts have been made by utility companies to improve power grid resilience under wildfire, the focus has been mostly on passive resilience techniques such as undergrounding power lines, vegetation management, pole reinforcing and other system hardening methods [4]. However, these passive methods centered on replacement, improvements, and hardening of electric infrastructure tend to be extremely costly. The costs associated with passive resilience can be reduced or deferred by incorporating automated control measures into system operation, an approach known as active (operational) resilience. Operational resilience can be categorized into three main actions: 1) preventive actions before wildfire events occur, 2) corrective actions after electrical systems hit by wildfires, and 3) restorative actions during system recovery.

### *A. Literature Review*

Recently, there has been growing interest in operational strategies for enhancing the resilience of electrical infrastructure in the face of extreme weather events. The study [5] proposed a multi-state transmission-system resilience framework for extreme weather events. The proposed model coordinates switching actions, generation rescheduling, and load shedding while accounting for uncertain transitions among weather conditions and the cascading impacts of short-circuit faults. Authors in [6] developed a corrective operation strategy that models the sequential evolution of system states during extreme events as a Markov process. Transition probabilities are evaluated based on event-induced failure rates, and the optimal strategy is obtained by minimizing a recursive value function combining current and future costs weighted by state probabilities. The study [7] proposes a resilience enhancement method combining corrective islanding with fast fault isolation and service restoration to improve distribution systems' resilience during extreme weather events. The interaction between power-system operation and wildfire ignition risk has also been investigated. In [8], Authors proposed an analytical framework to quantify the probability of wildfire ignition caused by distribution-line faults while accounting for environmental conditions, line loading, and protection-system settings..

In recent years, a growing body of research is examining how operational strategies can be leveraged to bolster the resilience of electrical infrastructure against wildfires. In this context, decision-making must consider the distinct spatiotemporal dynamics of wildfires and the probabilistic sequential failure of power system components. Authors in [9] propose a stochastic optimization strategy that enhances distribution grid resilience against progressive wildfires by modeling wildfire impacts on power lines and renewable generation, while jointly optimizing for resilience and equity objectives. In [10] authors propose a probabilistic generation redispatch strategy to enhance the operational resilience of power grids during wildfires. A Markov decision process is used to model system state transitions and determine redispatch actions based on component failure probabilities, wildfire dynamics, and load variations. Authors in [11] propose a mixed-integer quadratic optimization model to enhance the operational resilience of power distribution networks equipped with distributed energy resources under progressive wildfire conditions. The model optimally coordinates local resources to minimize load outages and improve system resilience. A decision support tool for managing power grids during wildfires is proposed in [12] such that various local generation resources (i.e., distributed renewable energy resources and energy storage systems) can be effectively employed to mitigate the wildfire impacts on the power grid. Authors in [13] propose a stochastic programming approach to determine the operation of a resilient distribution system exposed to an approaching wildfire. The study [14] proposes an optimization model to support short-term operational decision making for the public safety power shutoff (PSPS) problem in the context of extreme wildfire risk. In [15] the focus is on reclosing strategies that can be utilized by electric utilities for wildfire risk mitigation. In [16], the authors propose an optimization-based alternative to PSPS by introducing endogenous uncertainty into the decision process. Their model determines optimal network reconfigurations through switching actions to reduce power flow in vulnerable areas, thereby lowering wildfire ignition risk. The authors in [17] represent wildfire events as stochastic disruptions with uncertain magnitude and occurrence time. Their stochastic optimization framework seeks to maximize electricity delivery while strategically de-energizing system components across multiple time periods to reduce wildfire risk. A cellular automaton model is employed to simulate the initiation and spread of both exogenous and endogenous wildfires using environmental data.

### *B. Contributions*

The state-of-the-art models in the literature have simplified the operational resilience problem by ignoring the dependency between corrective strategies and preventive decisions (e.g., PSPS), and tackled these two problems separately. In practice, preventive actions like PSPS events impact the degree of freedom of making corrective decisions. That is, de-energizing electrical equipment such as power lines or energy resources reduces control variables and flexibility in the system for rerouting power, reconfiguring network, controlling energy resources and loads. As a result, these two decisions should be made in a consolidated multistage programming framework, in which the preventive decisions are made "here and now", and corrective decisions are made "wait and see" under uncertainty of wildfire spread scenarios.

Moreover, the mutual relationship between wildfires and electrical infrastructures complicates this problem even more, as both the uncertainty, probability distribution and the objective function in this problem depend on first stage decisions.

Wildfires exhibit distinctive characteristics compared to other extreme events. Unlike most hazards that solely impact the power grid, wildfires maintain a bidirectional relationship with electrical infrastructures, capable not only of damaging system components but also of being initiated by them. Consequently, the actions taken by system operators directly influence the risk of wildfires caused by electrical infrastructure. For example, implementing a PSPS changes line flows and the network topology, reducing ignition risk on lines where power flow decreases while increasing risk on lines where it rises. As a result, the spatial and temporal distribution of wildfire ignition risk is altered. Accordingly, the probability of wildfire spread and the resulting impact scenarios evolve based on preventive operational decisions. In this framework, corrective resilience actions are generally formulated according to the estimated likelihood of equipment damage and the prioritization of critical load supply during ongoing wildfire events. However, since PSPS operations can mitigate ignition risk in certain regions, they simultaneously reduce the likelihood of equipment damage within those areas. PSPS events may also cause electricity users to relocate to community resource centers or shelters situated outside the de-energized areas. Such population displacement elevates the priority of nearby shelter loads, thereby altering the priorities and operational constraints considered by system operators during wildfire events. Furthermore, PSPS decisions directly affect the feasible space of the corrective optimization problem by altering the network configuration. These are examples of how preventive decisions impact uncertainty, objective function, and constraints in corrective resilience optimization problems. Therefore, to obtain a more realistic and practical model, the impact of preventive decisions on objective function, constraints, and probability distribution of the stochastic parameters in optimization problems should be considered.

Most existing literature on grid resilience under wildfire conditions either focuses solely on exogenous uncertainties while overlooking endogenous (decision-dependent) uncertainties, or considers DDU only in long-term system planning, such as system hardening [18], line upgrades and resource allocation [19], and investment optimization including line upgrades and switching device placement [20]. However, these studies are limited to long-term planning and do not address real-time, multistage operational decision-making—particularly the interdependence between preventive PSPS decisions and corrective resilience actions. Furthermore, there remains a lack of comprehensive frameworks that systematically integrate both exogenous and endogenous uncertainties into multistage operational resilience scheduling.

The contribution of this paper can be summarized as

1) Establishing a multistage modeling framework as a foundation and extending it to capture the evolving mutual interaction between wildfire and electrical infrastructure, ensuring the problem formulation accurately reflects real-world conditions.
2) Developing and adapting integrated optimization frameworks to enhance resilience against wildfires, explicitly accounting for both endogenous (decision-dependent) and exogenous uncertainties.
3) Using advanced reformulation techniques and a novel, scalable, and computationally efficient automation algorithm to solve the proposed optimization model, which incorporates endogenous uncertainties into a multiage decision-making context.

The remainder of the paper is organized as follows: Section II presents the background and problem statement related to the multistage optimization framework and decision-dependent uncertainty (DDU) formulation. Section III outlines the key principles of the proposed solution algorithm. Section IV provides simulation results from applying the algorithm to the IEEE 30-bus system. Finally, Section V concludes the paper with a summary of the findings.

## II. Multistage Stochastic Program with Endogenous Uncertainty

In multistage stochastic optimization, DDU arises when decisions in one stage directly affect the evolution, realization, or probability distribution of uncertain parameters in subsequent stages. Existing literature classifies DDU into two classical types. Type I models the impact of decision variables on the timing at which uncertainty is (partially) revealed, capturing cases where decisions determine when information becomes available. Type II, on the other hand, represents situations where decisions influence the probability distribution of future events, thereby altering the scenario weights or likelihood of specific outcomes. However, in many complex systems—such as power grid operation under progressive wildfires—the interaction between decisions and uncertainty cannot be fully represented by either Type I or Type II. In wildfire-related optimization, preventive operational decisions not only affect the likelihood of future wildfire scenarios, particularly those initiated by power lines, but also alter the system configuration. These coupled effects simultaneously modify scenario probabilities, objective function coefficients, and constraint coefficients, leading to a richer and more intricate form of endogenous uncertainty. To address this, an extended structure [21], referred to by Type III DDU, has been proposed as a generalization of Type II. Type III models the influence of decision variables on both the probability distribution and the numerical coefficients that define the system's objectives and constraints within the scenario tree. This comprehensive formulation enables the stochastic model to capture the mutual interaction between decision making and the problem formulation, where preventive and corrective decisions reshape both the likelihood of wildfire scenarios and the operational and structural parameters of the system. Consequently, Type III DDU establishes a broader framework for representing complex decision–uncertainty interactions, allowing a more realistic characterization of wildfire evolution as a multistage stochastic process affected by adaptive decision-making.

This section presents the formulation of the problem under a Type III DDU framework, extending beyond the conventional multistage stochastic programming approach.

The problem of operational resilience of electrical systems against wildfire includes a sequence of decisions, which should be made under wildfire impact scenario uncertainty. Therefore, the optimization model should be formulated in the form of multistage stochastic programming to model the sequential nature of the process. To avoid confusion, whenever we refer to "nodes" in this paper, it specifically denotes the decision tree nodes, not the busbars of the power system. When discussing busbars, we will explicitly refer to them by their bus numbers to maintain clarity.

The state-of-the-art mathematical formulation of the preventive–corrective multistage problem for power system operation under wildfire conditions is provided in Appendix. In general, this formulation can be presented in (1)-(3).

$$\min of = \sum_{n\in\mathcal{N}} w^n \left(a^n x^n + b^n y^n\right) \tag{1}$$

$$s.t. \sum_{q\in\mathcal{A}^n} \left(A_n^q x^q + B_n^q y^q\right) = h^n \quad \forall n \in \mathcal{N} \tag{2}$$

$$x^n \in \{0,1\}^{nx^n}, \quad y^n \in \mathbb{R}^{ny^n} \quad \forall n \in \mathcal{N} \tag{3}$$

Let $w^n$ denote the weight associated with node $n \in \mathcal{N}$ that is computed as $w^n = \sum_{\omega\in\Omega^n} w^{\omega}$, where $\Omega^n$ represents the set of descendant scenarios of node *n*, and $\omega \in \Omega^n$. The decision vectors $x^n \in \{0,1\}^{nx^n}$ and $y^n \in \mathbb{R}^{ny^n}$ represent the binary and continuous variables, respectively, at node $n \in \mathcal{N}$. The vectors $a^n$ and $b^n$ contain the corresponding coefficients in the objective function associated with $x^n$ and $y^n$. The set $\mathcal{A}^n$ includes node *n* and its ancestors in the scenario tree whose variables have nonzero coefficients in the constraints at node *n*. For each node $n \in \mathcal{N}$, the matrices $A_n^q$ and $B_n^q$ represent the constraint coefficients, and $h^n$ denotes the right-hand side vector of the constraint system.

The general form of multistage mathematical formulation for preventive–corrective resilience strategies under wildfire conditions, presented in (1)-(3), should be reformulated to account for Type III DDU. This reformulation enables the explicit representation of the influence of first-stage preventive decisions on the uncertainty probability in subsequent stages, thereby capturing the dynamic interdependence between decision-making and uncertainty evolution in the system.

The reformulation of a multistage stochastic program with Type III decision-dependent uncertainties is express as (4)-(6):

$$\begin{aligned} min\, of = {} & a^1x^1 + b^1y^1 \\ & + \sum_{n\in\mathcal{N}\setminus 1} \sum_{k\in\Omega^n} p^k(x,y) \\ & \times \left[a^n(x,y).x^n + b^n(x,y).y^n\right] \end{aligned} \tag{4}$$

$$\begin{aligned} s.t. \sum_{q\in\mathcal{A}^n} & \left(A_n^q(x,y).x^q + B_n^q(x,y).y^q\right) \\ & = h^n(x,y), \forall\, n \in \mathcal{N} \end{aligned} \tag{5}$$

$$x^n \in \{0,1\}^{nx^n}, y^n \in \mathbb{R}^{ny^n}, \forall\, n \in \mathcal{N} \tag{6}$$

In this model, the probability of each scenario (i.e., $p^k(x,y)$), coefficients of variables $x^n$ and $y^n$ (i.e., $a^n(x,y)$ and $b^n(x,y)$), and constraint matrices (i.e., $A_n^q(x,y)$ and $B_n^q(x,y)$) are functions of decision variables.

To model and capture the impact of influential variables, a piecewise linear approximation is employed in place of nonlinear functions. In this context, the influential variables are a subset of decision variables, specifically first-stage decisions, that affect scenario probabilities, objective function coefficients, and constraint coefficients in subsequent stages.

By approximating the impact of decision variables in the form of piecewise linear functions, formulation (4)-(6) turns into a mixed integer bilinear model, as shown in (7)-(12).

$$\begin{aligned} min\, of = {} & a^1x^1 + b^1y^1 + d^1\sum_{\iota\in I^1} z_\iota^1 \\ & + \sum_{n\in\mathcal{N}\setminus\{1\}} \sum_{\iota\in I^1} \delta_\iota^1 \sum_{k\in\Omega^n} p_\iota^k \Bigg( a_\iota^n x^n \\ & + b_\iota^n y^n + d_\iota^n \sum_{\iota'\in I^n : t^n \le T-\tau} z_{\iota'}^n \Bigg) \end{aligned} \tag{7}$$

$$\begin{aligned} s.t. \sum_{q\in\mathcal{A}^n} & \left(A_n^q x^q + B_n^q y^q\right) + C^n \sum_{\iota\in I^n} z_\iota^n = h^n \\ & \forall\, n \in \mathcal{N}: t^n \le \tau \end{aligned} \tag{8}$$

$$\begin{aligned} \sum_{\iota\in I^\sigma} \delta_\iota^\sigma & \Bigg( \sum_{q\in\mathcal{A}^n} \left(A_{n,\iota}^q x^q + B_{n,\iota}^q y^q\right) \\ & + C_\iota^n \sum_{\iota'\in I^n : t^n \le T-\tau} z_{\iota'}^n \Bigg) \\ & = \sum_{\iota\in I^\sigma} \delta_\iota^\sigma h_\iota^n \\ & \forall\, n \in \mathcal{N}: t^n \ge \tau \end{aligned} \tag{9}$$

$$\begin{aligned} & \sum_{\iota\in I^\sigma} \delta_\iota^\sigma = 1 \\ & \forall\, \sigma \in \mathcal{N}: t^\sigma \le T-\tau \end{aligned} \tag{10}$$

$$\begin{aligned} & \bar{z}_{\iota-1}^\sigma \delta_\iota^\sigma \le z_\iota^\sigma \le \bar{z}_\iota^\sigma \delta_\iota^\sigma \\ & \delta_\iota^\sigma \in \{0,1\} \\ & \forall \iota \in \mathcal{I}^\sigma,\ \sigma \in \mathcal{N} : t^\sigma \le T-\tau \end{aligned} \tag{11}$$

$$\begin{aligned} & x^n \in \{0,1\}^{nx^n}, y^n \in \mathbb{R}^{ny^n} \\ & \forall\, n \in \mathcal{N} \end{aligned} \tag{12}$$

Equation (7) closely resembles the formulation in (4), with the primary distinction being the introduction of the influential variable $z_\iota^1$. This variable enables modifications to both the coefficients of objective function/constraint coefficients and the scenario probabilities associated with each node in the scenario tree. As a result, the objective function becomes a mixed-integer bilinear expression. Similarly, (8) differs from (5) for the same reason, while (9) illustrates how the influential variables can modify the coefficients of the nodal constraints. Constraints (10) and (11) ensure that only one disjoint value interval of $z$ is activated. Moreover, a latency parameter $\tau$ is incorporated into the formulation to specify that, prior to this time, the primary constraint coefficients remain unaffected by the influential variable $z$. This parameter models the predefined stage horizon within which influential variables do not yet affect the system constraints, representing the execution latency of operational actions.

Building upon the generic multistage optimization model with DDU, developed in (7)-(12), the preventive-corrective operational resilience problem in power systems is formulated within the DDU framework, as presented in (13)-(30). In this formulation, the introduction of an influential variable is required to capture the dependency between first-stage decisions and the evolution of uncertainty across future stages. Accordingly, wildfire ignition risk caused by electrical components, $R^{fire}$, is introduced as the influential variable and is defined in (16). By treating $R^{fire}$ as an influential variable, the impact of PSPS decisions on future scenario probabilities and objective function coefficients is effectively captured. The value of $R^{fire}$ dynamically reflects system-level operational decisions, including the energization or de-energization of transmission lines, generators, and busbars, as well as the percentage of load curtailment. These decisions influence the likelihood of wildfire ignition and propagation, thereby modifying the probability distribution of future scenarios and associated operational constraints. By defining $R^{fire}$as an influential variable, the model rigorously captures these decision-dependent interactions.

$$\begin{aligned} min\ [\alpha \sum_{\iota \in I^1} \left(R_\iota^{fire,1}\right) - ((1-\alpha)(D^{Tot,1} \\ - \sum_{g \in G} C^{g,1}(P^{g,1})) \\ + \sum_{n \in N \backslash \{1\}} \sum_{\iota \in I^1} \delta_\iota^1 \omega_\iota^1 ((1 \\ - \alpha)(\sum_{g \in G} C_\iota^{g,n}(P^{g,n}) - D_\iota^{Tot,n}) \\ + \alpha \sum_{\iota' \in I^1; t^n \leq T-\tau} (R_{\iota'}^{fire,n})] \end{aligned} \tag{13}$$

$$\sum_{\iota \in I} \delta_\iota^\sigma = 1 \quad \forall\, \sigma \in \mathcal{N}: t^\sigma \leq T - \tau \tag{14}$$

$$\bar{R}_\iota^{fire,\sigma-1} \delta_\iota^\sigma \leq R_\iota^{fire,\sigma} \leq \bar{R}_\iota^{fire,\sigma} \delta_\iota^\sigma \quad \forall \iota \in I^\sigma,\ \ \sigma \in \mathcal{N}:\ t^\sigma \leq T - \tau \tag{15}$$

In this regard, the objective function of the DDU-based optimization model is expressed in (13), while (14) corresponds directly to the generic formulation defined in (10). Moreover, selecting $R^{fire}$ as the influential variable, (11) is modified and reformulated as (15) which is constructed based on the piecewise linear weighted aggregation of $R^{fire}$.

This formulation seeks to minimize the expected wildfire risk $R^{fire,n}$, maximize the weighted total delivered load $D^{Tot,n}$, and minimize the generation operational cost $C^{g,n}(P^{g,n})$ across all nodes $n \in N$. The parameter $\alpha \in [0\ 1]$ governs the trade-off among these competing objectives, balancing the priorities of wildfire risk mitigation, load delivery, and cost efficiency. The proposed DDU-embedded formulation enables the impact of the influential variable to propagate throughout the multistage decision process.

$$\begin{aligned} R^{fire,n} = \sum_{d \in D} l^{d,n} R^{d,n} + \sum_{g \in G} u^{g,n} R^{g,n} \\ + \sum_{e \in E} u^{e,n} R^{e,n} + \sum_{i \in \beta} u^{i,n} R^{i,n} \\ t_n \leq \tau \end{aligned} \tag{16}$$

$$D^{Tot,n} = \sum_{d \in D} l^{d,n} \psi^{d,n} D^{d,n} \quad t_n \leq \tau \tag{17}$$

$$u^{i,n} \geq l^{d,n} \quad \forall d \in \beta_i^D\ \forall i \in \beta, t_n \leq \tau \tag{18}$$

$$u^{i,n} \geq u^{g,n} \quad \forall g \in \beta_i^G\ \forall i \in \beta, t_n \leq \tau \tag{19}$$

$$u^{i,n} \geq u^{e,n} \quad \forall e \in \beta_i^e\ \forall i \in \beta, t_n \leq \tau \tag{20}$$

$$P^{e,n} \leq -b^{e,n}(\theta^{i,n} - \theta^{j,n} + \theta^{max}(1 - u^{e,n})) \quad \forall i \in \beta, \forall j \in \beta, e \in E(i,j), t_n \leq \tau \tag{21}$$

$$P^{e,n} \geq -b^{e,n}(\theta^{i,n} - \theta^{j,n} + \theta^{min}(1 - u^{e,n})) \quad \forall i \in \beta, \forall j \in \beta, e \in E(i,j), t_n \leq \tau \tag{22}$$

$$-T^e u^{e,n} \leq P^{e,n} \leq T^e u^{e,n} \quad t_n \leq \tau \tag{23}$$

$$\sum_{g \in \beta_i^G} P^{g,n} = \sum_{e \in \beta_i^e} P^{e,n} + \sum_{d \in \beta_i^d} l^{d,n} D^{d,n} \quad \forall i \in \beta, e \in E(i,j), t_n \leq \tau \tag{24}$$

$$P^{g,n} = P^{g,a(n)} + \Delta P^{g,n} \quad \forall g \in G, t_n \leq \tau \tag{25}$$

$$\Delta p_{min}^g \leq \Delta P^{g,n} \leq \Delta p_{max}^g \quad \forall g \in G, t_n \leq \tau \tag{26}$$

$$u^{g,n} P_{min}^g \leq P^{g,n} \leq u^{g,n} P_{max}^g \quad \forall g \in G, t_n \leq \tau \tag{27}$$

In the DDU-based multistage formulation, all constraints originally defined in the standard multistage formulation are required to hold for all time steps $t_n \leq \tau$, as prescribed in (8). Accordingly, (16)-(27) present the set of multistage constraints corresponding to these time steps, which must be satisfied within the DDU optimization framework. The description of constraints (16)-(27) is provided in Appendix.

For later stages where $t_n > \tau, t_n \leq T - \tau$, additional decision-dependent constraints must be defined, as specified in (9), to capture the influence of influential decision variables on the affected constraints. To this end, the parameter $\delta_\iota^\sigma$ is incorporated into affected constraint equations to ensure consistency with the piecewise linear reformulation adopted in the DDU framework. This integration enables the constraints to accurately capture the impact of the influential variables on system behavior and align the mathematical formulation with the decision-dependent uncertainty representation. Specifically, in (28), $\delta_\iota^\sigma$ links the wildfire risk to the piecewise linear function, allowing the model to represent how preventive decisions influence the probability of future scenarios over the multistage time horizon of the wildfire. The term $R_\iota^{fire,n}$ represents the piecewise linear, node-specific wildfire ignition risk caused by electrical components at node $n$, which varies at different stages and scenarios based on power flow conditions and environmental factors. Equation (29) illustrates how preventive decisions based on $R^{fire}$ can influence future objective function coefficients. Depending on these decisions and the evolution of future scenarios, the priority and demand of critical loads- such as shelters- can be adjusted. Here, $\psi_\iota^{d,n}$ represents a weighting factor reflecting the prioritization of load $d$, while $D_\iota^{d,n}$ denotes the corresponding demand, which are dependent on the preventive decisions using a piecewise linear function. The nodal power balance relation is updated in (30) to reflect the impact of influential variables on optimization constraints through $\delta_\iota^\sigma$.

Since the preventive decisions do not affect the coefficients of constraints (18)-(23) and (25)-(27), these constraints remain unchanged for $t_n > \tau$ and $t_n \le T - \tau$. Across all reformulated constraints, the parameter $\delta_\iota^\sigma$ is employed to construct lexicographically ordered disjoint value sets, ensuring a consistent piecewise linear representation within the DDU framework. This formulation embeds the decision-dependent uncertainties directly into the system's operational dynamics through the state of the influential variable $R^{fire}$.

$$\sum_{\iota\in I^\sigma} \delta_\iota^\sigma R_\iota^{fire,n} = \sum_{\iota\in I^\sigma} \delta_\iota^\sigma (\sum_{d\in D} l^{d,n} R_\iota^{d,n} + \sum_{g\in G} u^{g,n} R_\iota^{g,n} + \sum_{e\in E} u^{e,n} R_\iota^{e,n} + \sum_{i\in\beta} u^{i,n} R_\iota^{i,n}) \tag{28}$$

$$\iota \in I^\sigma, \forall n \in N, t_n > \tau, t_n \le T - \tau$$

$$\sum_{\iota\in I^\sigma} \delta_\iota^\sigma D_\iota^{Tot,n} = \sum_{\iota\in I^\sigma} \delta_\iota^\sigma \sum_{d\in D} l^{d,n} \psi_\iota^{d,n} D_\iota^{d,n} \tag{29}$$

$$\iota \in I^\sigma, \forall n \in N, t_n > \tau, t_n \le T - \tau$$

$$\sum_{\iota\in I^\sigma} \delta_\iota^\sigma \sum_{g\in\beta_i^G} P^{g,n} = \sum_{\iota\in I^\sigma} \delta_\iota^\sigma (\sum_{e\in\beta_i^e} P^{e,n} + \sum_{d\in\beta_i^d} l^{d,n} D_\iota^{d,n}) \tag{30}$$

$$\forall i \in \beta, e \in E(i,j), \iota \in I^\sigma, \forall n \in N, t_n > \tau, t_n \le T - \tau$$

s.t.: (18)-(23)

(25)-(27)

$$t_n > \tau, t_n \le T - \tau$$

## III. Solution Algorithm

In a large-scale power system, the optimization problem (13)-(30) cannot be solved within a reasonable timeframe due to the presence of mixed-integer bilinear terms in the model, as well as its stochastic multistage structure. In this paper, the Regularized Cluster Progressive Algorithm (RCPA) is proposed to solve each cluster of the optimization problem introduced in the previous section. RCPA extends the classical Progressive Hedging Algorithm by incorporating scenario clustering and regularization to improve convergence and solution quality in stochastic mixed-integer settings. At each iteration, the algorithm solves the subproblem associated with each scenario cluster, updates the pseudo-gradient terms in the objective functions, and computes an implementable solution by averaging the decisions across clusters. This averaged solution $(\hat{X}_T, \hat{Y}_T)$ is then used as a reference in subsequent iterations, guiding the search toward a feasible and high-quality solution for the original problem [22], [23]. Algorithm 1 provides solution procedure for the proposed multistage DDU optimization problem.

Step 1 of Algorithm 1 provides an initial non-implementable solution by solving each scenario-cluster MILP independently. This produces the first set of decisions for all scenario clusters $s \in S$, which serve as the starting point for the iterative coordination process. To compute the pseudo-gradient scaling parameters used in Step 2, the algorithm first determines the maximum and minimum values that each binary variable takes across all scenario clusters, as expressed in (31).

$$(\hat{X}_J)_{T_1}^{max} := argmax_{s\in S}(\hat{X}_J)_{T_{1s}}, (\hat{X}_J)_{T_1}^{min} := argmax_{s\in S}(\hat{X}_J)_{T_{1s}} \quad for\ J = 1, \dots, nx + n\delta. \tag{31}$$

Using these ranges, the algorithm computes a scaling coefficient $\rho^X$ for each binary variable, which determines the strength of the pseudo-gradient adjustment applied in Step 2. This scaling ensures that variables exhibiting greater dispersion across scenario clusters receive stronger coordination penalties. The resulting vector $\rho^X$, associated with the binary variable set $X_{T_1}$, consists of scalar components $\{(\rho^X)_j : j = 1, \dots, n_x + n_\delta\}$, each computed according to (32).

$$(\rho^X)_J := \frac{\Gamma \times (a_J)_{T_1}}{(\hat{X}_J)_{T_1}^{max} - (\hat{X}_J)_{T_1}^{min} + 1} \tag{32}$$

Here, $\Gamma$ is a positive scalar multiplier that determines the emphasis placed on the objective function coefficient $(a_J)_{\Gamma_1}$, specifically, larger values of $\Gamma$ indicate a higher weighting or prioritization of coefficient $a_J$ in the optimization process.

The parameter vector $\rho^Y$, corresponding to the continuous vector $Y_{T_1}$, comprises a set of scalar elements $\{(\rho^Y)_J$ for $J = 1, \dots, ny + nz\}$, where each scalar $(\rho^Y)_J$ is determined according to (33).

$$(\rho^Y)_J := \frac{\Gamma \times (b_J)_{T_1}}{max\left\{\sum_{s\in S} w_S \left|(Y_J)_{T_{1s}} - (\hat{Y}_J)_{T_1}\right|, 1\right\}} \tag{33}$$

It is important to recognize that the denominator in (33) represents the expected $l_1$-norm deviation of each continuous variable from its current implementable value across all scenario clusters. This term normalizes the scaling factor so that variables with larger average deviations receive proportionally stronger penalization.

With these scaling parameters in place, Step 2 updates the pseudo-gradient terms for all scenario clusters using the expressions in (34), which increases the penalty for deviations from the implementable solution and promotes convergence toward a coordinated set of decisions.

$$\mu_s^X := \mu_s^X + \rho^X (X_{T_{1s}} - X_{T_1}) and \quad \mu_s^Y := \mu_s^Y + \rho^Y (Y_{T_{1s}} - Y_{T_1}) \tag{34}$$

In Step 3, the Lagrangian bound corresponding to each scenario cluster indexed by $s \in S$ is computed by solving the generic MILP model in (7)-(12). Within this step, the original objective function term $a_{T_1} X_{T_{1s}} + b_{T_1} Y_{T_{1s}}$ is replaced with the modified expression $(a_{T_1} + \mu_s^X) X_{T_{1s}} + (b_{T_1} + \mu_s^Y) Y_{T_{1s}}$. As a result, the objective function in model (7)-(12) is substituted with the updated formulation given in (35).

$$of_{\mu_s} = of_s + \mu_s^X X_{T_{1s}} + \mu_s^Y Y_{T_{1s}} \tag{35}$$

In Algorithm 1, the quantity $\widehat{of_s}$ corresponds to the value of the modified objective $of_{\mu_s}$ obtained after solving the MILP for cluster $s$. This value is then used to update the Lagrangian bound of the original model.

In Step 4, each scenario-cluster subproblem is solved again, this time including a regularization term that penalizes the deviation of the local scenario-specific decisions from the current implementable solution. For every scenario $s \in S$, the model introduces an $l_1$-norm regularization that drives the binary variables $X_{T_{1s}}$ and continuous variables $Y_{T_{1s}}$ toward their implementable counterparts $X_{T_1}$ and $Y_{T_1}$. The deviation of the continuous variables from their implementable values is handled by introducing two non-negative auxiliary vectors $Y_{+,T_{1s}}$ and $Y_{-,T_{1s}}$ which satisfy the decomposition $Y_{T_{1s}} - \hat{Y}_{T_1} = Y_{+,T_{1s}} - Y_{-,T_{1s}}$. The resulting MILP solved in this step corresponds exactly to model (36) in Algorithm 1, where the objective combines: (i) the original objective value $of_s$, (ii) the pseudo-gradient terms $\mu_s^X$ and $\mu_s^Y$, and (iii) the quadratic-regularization contributions $\rho^X$ and $\rho^Y$. Solving this updated subproblem yields the new non-implementable solution $(\hat{X}_{T_{1s}}, \hat{Y}_{T_{1s}})$ for each scenario cluster, which will be used in Step 5 to update the implementable solution.

$$\min_{of_1,\mu_s} \frac{\rho^X}{2}\left\|\hat{X}_{T_1}\right\| + of_s + \left(\mu_s^X + \frac{\rho^X}{2} - \rho^X \hat{X}_{T_1}\right) X_{T_{1s}} + \left(\mu_s^Y - \rho^Y \hat{Y}_{T_1}\right) Y_{T_{1s}} + \frac{\rho^Y}{2}\left(Y_{+,T_{1s}} - Y_{-,T_{1s}}\right) \tag{36}$$

s.t. the constraint system of model (7)-(12) and
$Y_{T_{1s}} - \hat{Y}_{T_1} = Y_{+,T_{1s}} - Y_{-,T_{1s}}, \quad Y_{+,T_{1s}}, Y_{-,T_{1s}} \in \mathbb{R}^+$

**Algorithm 1:** ***Solution procedure for multistage DDU optimization.***

1: **(First (likely) non-implementable solution)**
Solve each of the S scenario cluster MILP models (7)-(12).
Output: $(\hat{X}_s, \hat{Y}_s, \widehat{of}_s, \forall s \in S)$.
If the solution $(\hat{X}_{T_{1s}}, \hat{Y}_{T_{1s}}\ \forall s \in S)$ is implementable: set $\overline{ob}_{ENDO} := \sum_{s\in S} \widehat{of}_s$ and stop1.
Set first Lagrangian bound: $\underline{ob}_{ENDO} := \sum_{s\in S} \widehat{of}_s$.
Initialize parameters: $k := 1, \mu_s^X := \mu_s^Y := 0\ \forall s \in S, (\hat{X}_{T_1}, \hat{Y}_{T_1}) := \sum_{s\in S} w_s(\hat{X}_{T_{1s}}, \hat{Y}_{T_{1s}})$

2: **Weighted non-implementable solution and pseudo-gradient computations**
Modify the current solution: $(\hat{X}_{T_{1s}}, \hat{Y}_{T_{1s}}) := (1-\alpha)(\hat{X}_{T_1}, \hat{Y}_{T_1}) + \alpha(\hat{X}_{T_{1s}}, \hat{Y}_{T_{1s}})\ \forall s \in S$
Reset pseudo-gradients $\mu_s^X$ and $\mu_s^Y$ using (34)

3: **Lagrangian bound updating**
Solve MILP model (7)-(12) for scenario cluster indexed with s, replacing the objective function with function (35).
Output: $\widehat{of}_s$.
Update $\underline{ob}_{ENDO} := \max\{\sum_{s\in S} \widehat{of}_s, \underline{ob}_{ENDO}\}$

4: **Non-implementable solution retrieval**
Solve cluster model (36).
Output $(\hat{X}_s, \hat{Y}_s, \widehat{of}_s)$.

5: **Implementable solution resetting**
Reset $(\hat{X}_{T_1}, \hat{Y}_{T_1}) := \sum_{s\in S} w_s(\hat{X}_{T_{1s}}, \hat{Y}_{T_{1s}})$.
Check convergence conditions:
If $\sum_{s\in S} w_s\left\|\hat{X}_{T_{1s}} - \hat{X}_{T_1}\right\|^2 \le \epsilon_1$, $\hat{X}_{T_1}$ is an all {0,1} value.
If $\sum_{s\in S} w_s\left\|\hat{Y}_{T_{1s}} - \hat{Y}_{T_1}\right\|^2 \le \epsilon_2\left\|\hat{Y}_{T_1}\right\|^2$, then set $\overline{ob}_{ENDO} := \sum_{s\in S} \widehat{of}_s$ and stop 2.

6: **Testing a stopping criterion**
If $k < \bar{k},\ k := k+1$ and go to Step 2.
Execute a rounding and fixing scheme: $(X_J)_{T_1} := \lfloor(\hat{X}_J)_{T_1} + 0.5\rfloor$
Solve the original model ENDO (7)-(12) $(\hat{X}, \hat{Y}, \overline{ob}_{ENDO})$.
Compute optimality $gapOG\% := 100.\frac{\overline{ob}_{ENDO} - \underline{ob}_{ENDO}}{\underline{ob}_{ENDO}}$ and stop 3.

In Step 5, the algorithm updates the implementable solution by recomputing $(\hat{X}_{T_1}, \hat{Y}_{T_1})$ as the probability-weighted average of the non-implementable solutions $(\hat{X}_{T_{1s}}, \hat{Y}_{T_{1s}})$ obtained in Step 4 for all scenario clusters. After resetting $(\hat{X}_{T_1}, \hat{Y}_{T_1})$, the algorithm immediately checks the convergence conditions.

If the maximum number of iterations is reached, Step 6 performs a final attempt to obtain a feasible solution for the original model. In this step, a {0,1} implementable solution $(X_J)_{T_1}$ is generated from the current non-implementable [0,1] solution through a rounding procedure defined as $(X_J)_{T_1} := \lfloor(\hat{X}_J)_{T_1} + 0.5\rfloor\ \forall J = 1, \dots, nX + n\delta$. Subsequently, model (7)-(12) is solved with the corresponding variables set to these rounded values. In cases where the solution cannot be obtained under the given conditions, this approach selects the most frequently occurring approximate {0,1} value for these variables, effectively incorporating a form of a lazy algorithm within the solution scheme.

## IV. Simulation and Result

The effectiveness of the proposed methodology is demonstrated through its application to the IEEE 30-bus test system. The corresponding power generation data is presented in TABLE I.

In Figure 1, a wildfire-prone region within the IEEE 30-bus system is identified, where elevated ignition risk due to vegetation conditions and the deteriorated state of power line infrastructure, necessitates PSPS decisions. In this context, three transmission lines including lines 6–10, 6–4, and 10–17, are considered to have a high potential for wildfire ignition, and thus are subject to PSPS evaluation.

**TABLE I** Generation Data

| Unit | Power (MW) | | Cost (\$/MW) |
|---|---|---|---|
| | Min | Max | |
| $G_1$ | 50 | 200 | 20 |
| $G_2$ | 20 | 80 | 17.5 |
| $G_3$ | 15 | 50 | 10 |
| $G_4$ | 10 | 35 | 32.5 |
| $G_5$ | 10 | 30 | 30 |
| $G_6$ | 12 | 40 | 30 |

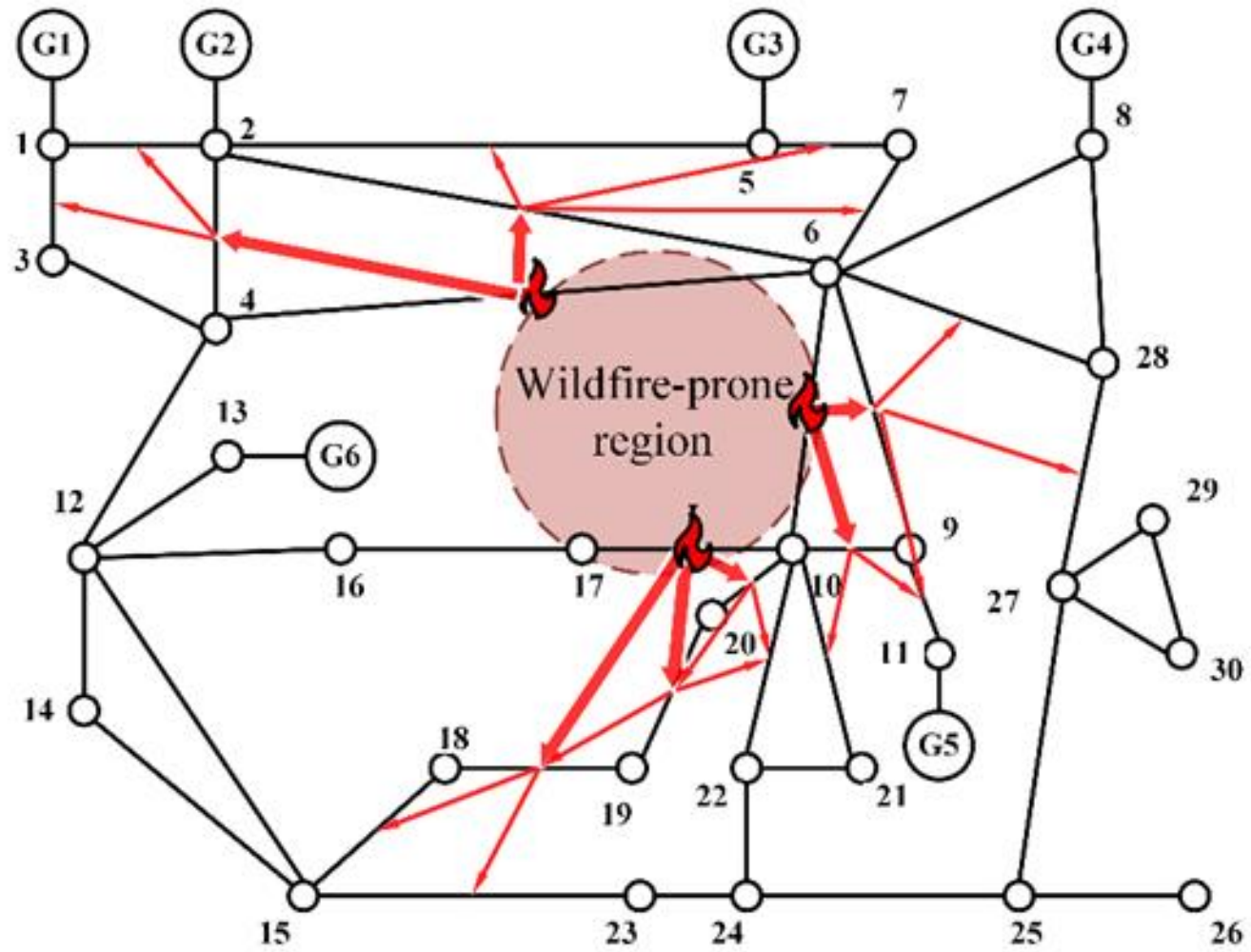

**Figure 1** IEEE 30-bus system highlighting the wildfire-prone region and associated wildfire propagation scenarios.

In conventional approaches, PSPS decisions are typically made based solely on the individual ignition risk of each component. However, in the proposed methodology, it is acknowledged that PSPS actions can alter the overall line flow within the network, subsequently affecting the probabilities of ignition caused by other components. As such, the probability of wildfire ignition by electrical infrastructure becomes decision-dependent, meaning that the operator's preventive or corrective actions directly influence how the uncertainty evolves over time. To represent this interdependence, the model adopts a DDU-based multistage optimization framework, in which future scenarios are generated according to potential wildfire spread paths and their spatial progression toward system components, while explicitly accounting for the impact of current decisions on the probability of future scenarios and associated constraints.

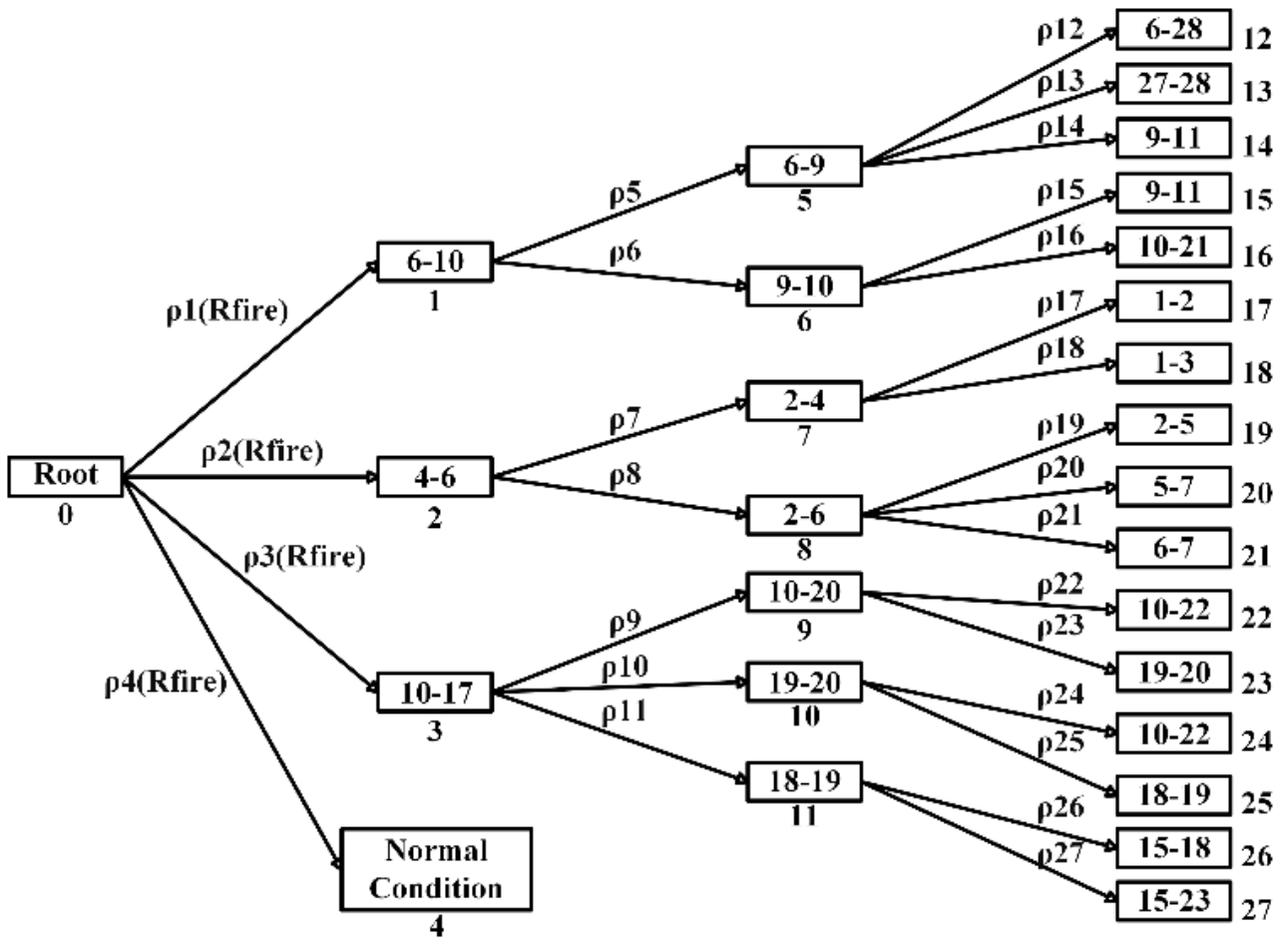

**Figure 2** Multistage DDU scenario tree representing wildfire ignition and propagation scenarios in the IEEE 30-bus system.

For each of the three identified high-risk transmission lines, three distinct wildfires spread scenarios are modeled based on their corresponding risk levels. Additionally, a normal operation scenario (i.e., the absence of wildfire events) is incorporated into the scenario tree, as illustrated in Figure 2. Within each scenario, the probability distribution of wildfire occurrence is dynamically updated based on the decisions made in preceding stages, reflecting the decision-dependent nature of the uncertainty. The probability distribution is updated based on influential variables -for example, whether a line is energized or de-energized- that determine the likelihood of wildfire ignition and spread in subsequent stages.

When a line is proactively de-energized through a PSPS decision, its associated ignition probability is reduced to zero. However, this action alters power flow, which may, in turn, increase or decrease the ignition risk of other lines in the wildfire-prone region. As a result, the probability of a normal operation scenario is also modified.

As previously discussed, these PSPS decisions, made in the initial stage, influence the probability distribution at the next stage of the multistage DDU model. Furthermore, depending on future scenarios, the prioritization of specific loads, such as emergency shelters, may change based on implemented PSPS actions, thereby impacting the value of serving these loads and affecting the overall optimization and decision-making process.

The proposed multistage DDU optimization framework is applied to this use-case, and the results are compared with those obtained from a model that does not account for the impact of decision-dependent uncertainty.

Different ignition risk levels are assigned to transmission lines within the wildfire-prone region, primarily based on the power flow through those lines. In the DDU-based scenario framework, whenever the optimization model determines that a line should be de-energized as part of a PSPS decision, the corresponding ignition risk, and consequently the probability of future wildfire-related scenarios, changes. This dependency is explicitly modeled for three high-risk lines. Additionally, in each future scenario, the value associated with shelter loads varies depending on whether a PSPS action was taken and whether a wildfire propagates through the affected area.

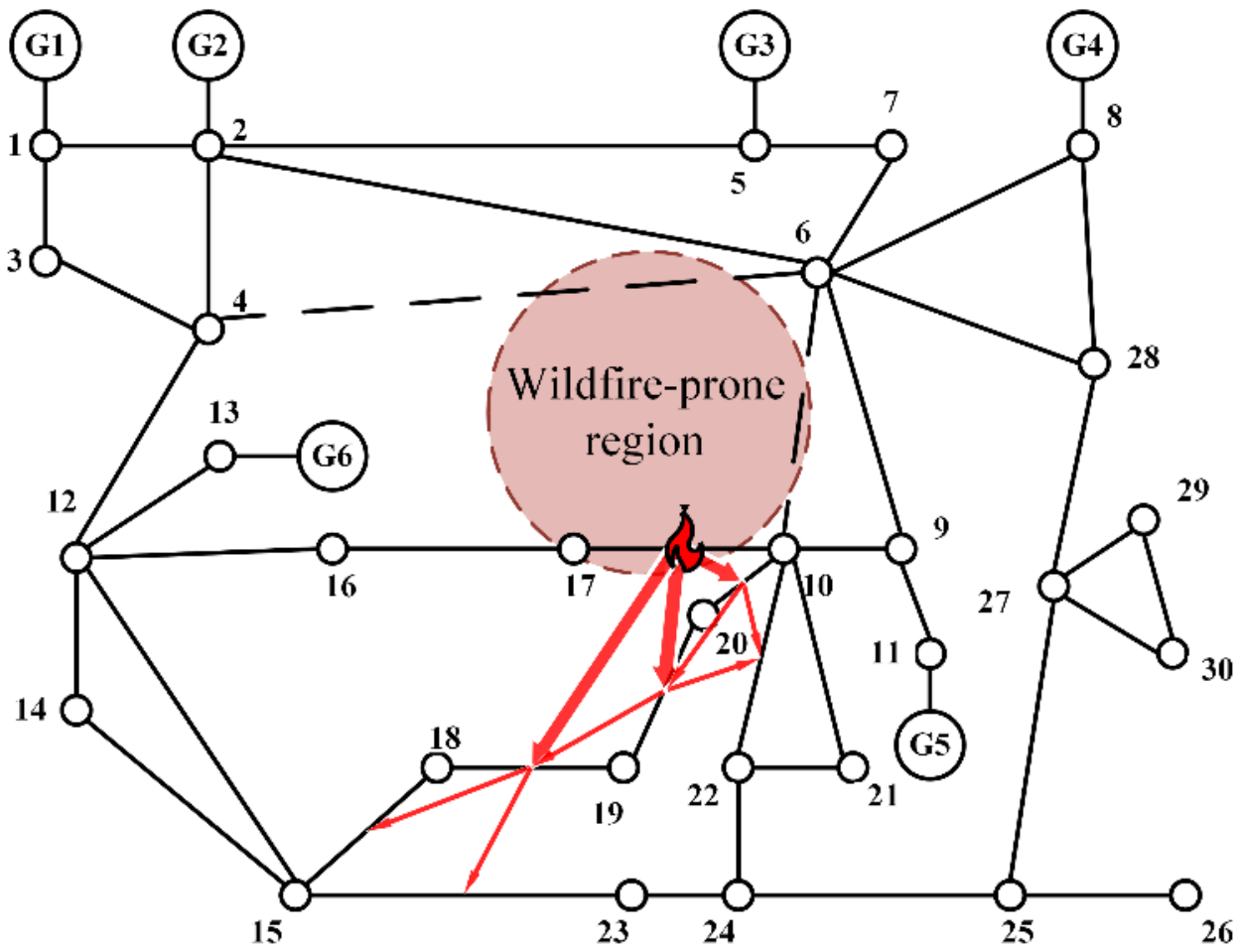

**Figure 3** Wildfire propagation path following the PSPS decision to de-energize lines 6-10 and 4-6.

By applying the multistage DDU optimization framework to the case study, the model identifies lines 6–10 and 4-6 for proactive de-energization. As illustrated in Figure 3, this decision effectively reduces the ignition probability along these paths to zero. However, this also alters the probability distribution, including the probabilities of normal operation scenario and other wildfire spread scenarios (i.e., scenarios associated with ignition caused by 10-17).

To validate the effectiveness of the DDU-based approach, the same operational setup is analyzed using a conventional multistage optimization model that does not account for decision-dependent uncertainty. As a result of this optimization, lines 6–10 and 10–17 were chosen for de-energization. A comparative analysis between the two approaches reveals that the DDU-based model results in a lower cost associated with operations, risk, and load loss, and more efficient decision-making.

From Stage 1 to Stage 2, both models experience a cost reduction due to improved corrective dispatch under the updated network topology and the elimination of ignition risk on de-energized lines. However, from Stage 2 onward, the models diverge. The non-DDU model shows a significant cost increase in Stages 3 and 4 due to unintended power flow redistribution onto line 4–6, increasing its ignition risk. In contrast, the DDU model anticipates these effects and maintains more stable costs. Overall, the DDU approach achieves approximately a 7% lower expected cost, demonstrating the value of modeling decision-dependent uncertainty.

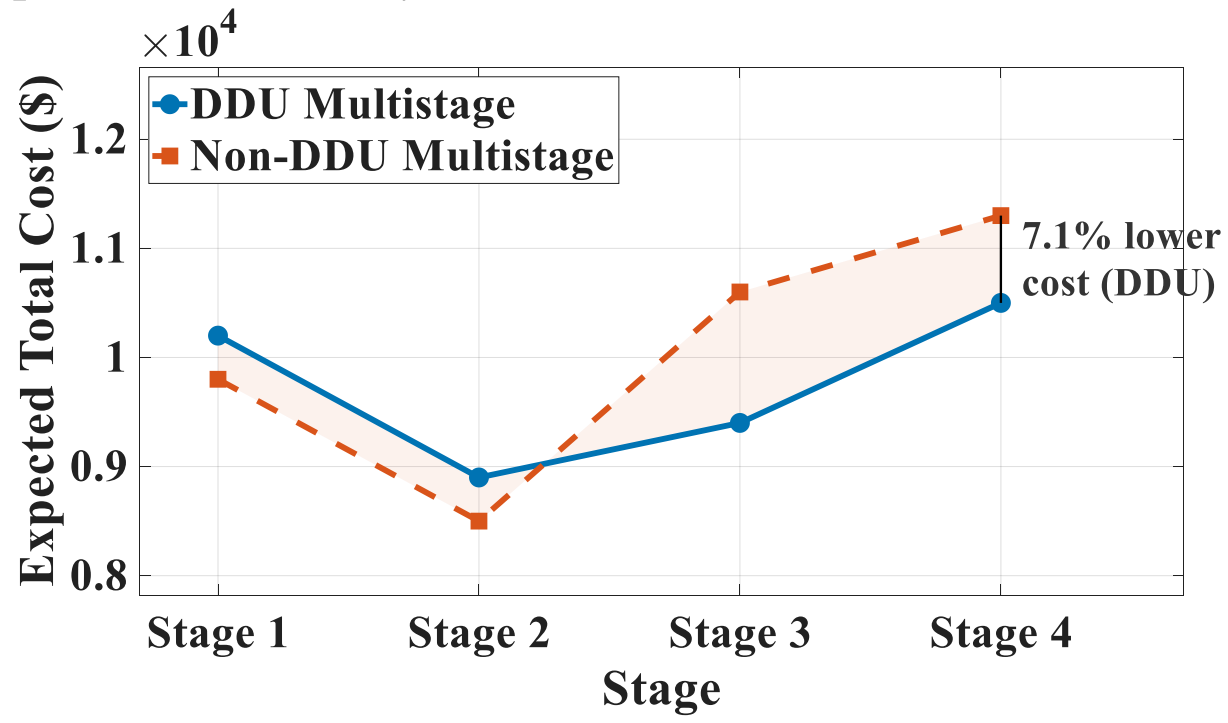

**Figure 4** Stage-wise expected total cost comparison between the proposed DDU multistage model and the conventional non-DDU multistage model on the IEEE 30-bus system.

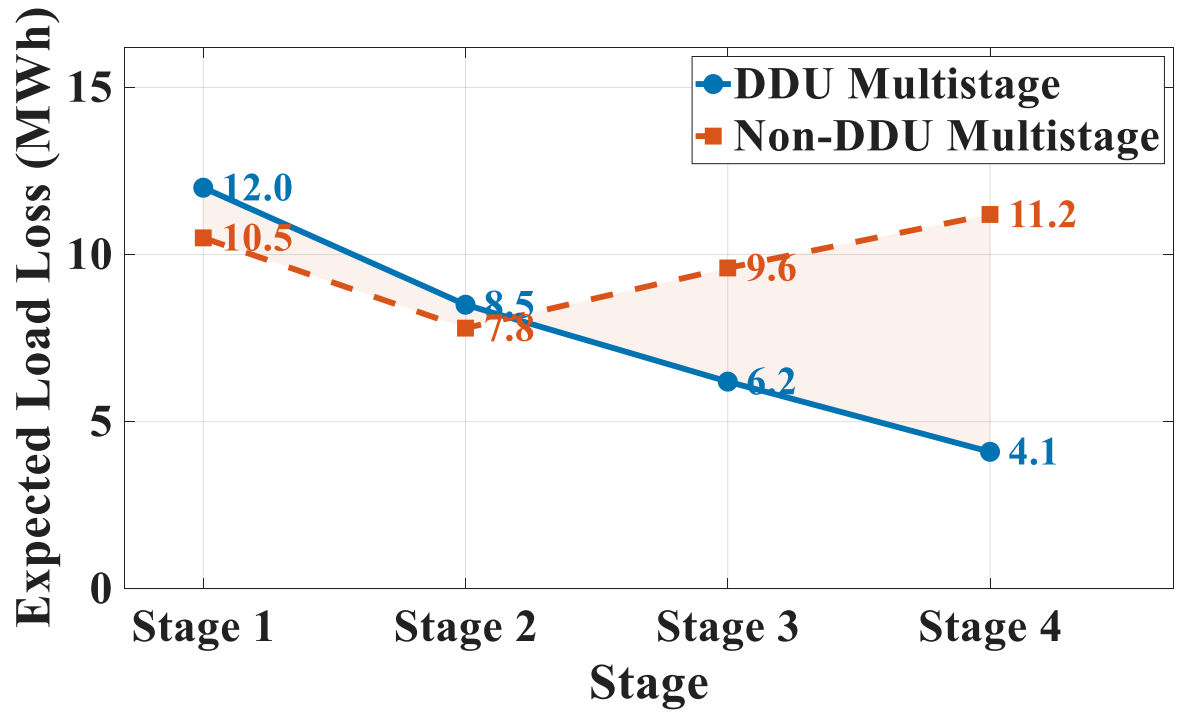

**Figure 5** Stage-wise expected load loss (MWh) for the proposed DDU multistage model and the conventional non-DDU multistage model on the IEEE 30-bus system.

Figure 5 presents the stage-wise expected load loss (MWh), reflecting the impact of PSPS and corrective actions on service continuity. At Stage 1, the non-DDU model exhibits slightly lower load loss (10.5 MWh vs. 12 MWh), as de-energizing line 10–17 affects a smaller immediate load compared to line 4–6. However, this short-term advantage diminishes in subsequent stages.

The DDU model achieves a steady reduction in load loss from 10.5 MWh at Stage 1 to 4.1 MWh at Stage 4, driven by improved network flexibility resulting from a more informed preventive decision. As corrective actions are implemented, generation redispatch and power rerouting enable progressive load restoration. In contrast, the non-DDU model experiences increasing load loss from Stage 2 onward, rising from 7.8 MWh to 11.2 MWh at Stage 4. This trend is caused by unanticipated power flow increases on line 4–6, pushing it toward thermal limits and necessitating additional emergency curtailment in later stages. The DDU model avoids this outcome by proactively de-energizing line 4–6 at Stage 1, thereby enabling more efficient load recovery across subsequent stages.

## V. Conclusion

In conclusion, this paper developed a comprehensive multistage optimization framework that incorporates decision-dependent uncertainty (DDU) to enhance the operational resilience of power systems against wildfires. By modeling both preventive and corrective actions and capturing the relationship between PSPS decisions and the future risk scenarios, the proposed DDU-based approach enables more informed and adaptive decision-making over time. Importantly, the model accounts for how early-stage decisions influence not only the probability distributions of future scenarios but also the key parameters that shape subsequent optimization stages. The framework was validated using the IEEE 30-bus test system, demonstrating its effectiveness in reducing risks and operational costs as well as improving system reliability compared to conventional models that ignore such dependencies. The results highlight the critical importance of incorporating the interdependence between operational decisions and evolving uncertainties in both scenario modeling and optimization processes, particularly under complex and progressive threats such as wildfires.

## Appendix

This section presents a preventive-corrective multistage optimization model for power systems under wildfire conditions. the objective function (A1) aims to minimize the expected wildfire risk $R^{fire,n}$, maximize the weighted load delivery $D^{Tot,n}$, and minimize generation costs $C^{g,n}(P^{g,n})$ for each node $n \in N$. The parameter $\alpha \in [0\ 1]$ controls the trade-off between wildfire risks, load delivery, and generation cost. The expected value formulation enables a thorough evaluation of objective function across all stages and nodes (i.e., scenarios) within the multistage optimization framework.

$$min\ \mathbb{E}[\sum_{n\in N}(\alpha R^{fire,n} - (1-\alpha)(D^{Tot,n} - \sum_{g\in G} C^{g,n}(P^{g,n}))] \tag{A1}$$

The extended nodal formulation of (A1) is presented in (A2), where $\rho^n$ denotes the probability associated with the occurrence of each scenario at a specific node within the decision tree.

$$min\ [\sum_{n\in N}(\rho^n(\alpha R^{fire,n} - (1-\alpha)(D^{Tot,n} - \sum_{g\in G} C^{g,n}(P^{g,n}))] \tag{A2}$$

The optimization model uses binary variables $u^{*,n} \in \{0,1\}$, for the energization status of generators, lines, and buses, where a value of 1 indicates that the component is energized, and 0 indicates it is de-energized at node $n$. Specifically, the variables $u^{g,n}, u^{e,n}$ and $u^{i,n}$ represent the status of generators, lines, and buses, respectively. Load shedding is represented by a continuous decision variable $l^{d,n} \in [0\ \ 1]$, which denotes the proportion of the load that is curtailed at node $n$. In this formulation, the ignition risk associated with electrical components at node $n$ is represented by $R^{fire,n}$ , as defined in (A3). This node-specific risk metric captures the stage-dependent nature of wildfire risk and its variability across components, influenced by power flow conditions and environmental factors. The total weighted load delivered at node $n$, $D^{Tot,n}$, is defined in (A4). In this expression, $D^{d,n}$ represents the load $d$ at node $n$, while $\psi^{d,n}$ is a weighting factor that reflects the prioritization of load $d$ during wildfire events. Dependencies of the status of generators, loads, and lines on bus energization status are enforced through (A5)-(A7).

$$R^{fire,n} = \sum_{d\in D} l^{d,n}R^{d,n} + \sum_{g\in G} u^{g,n}R^{g,n} + \sum_{e\in E} u^{e,n}R^{e,n} + \sum_{i\in B} u^{i,n}R^{i,n} \tag{A3}$$

$$D^{Tot,n} = \sum_{d \in D} l^{d,n} \psi^{d,n} D^{d,n} \quad \text{(A4)}$$

$$u^{i,n} \geq l^{d,n} \quad \forall d \in \beta_i^D \; \forall i \in \beta \quad \text{(A5)}$$

$$u^{i,n} \geq u^{g,n} \quad \forall g \in \beta_i^G \; \forall i \in \beta \quad \text{(A6)}$$

$$u^{i,n} \geq u^{e,n} \quad \forall e \in \beta_i^e \; \forall i \in \beta \quad \text{(A7)}$$

Equations (A8) to (A10) model DC power flow, with line characteristics defined by thermal limits $T^e$ and susceptance $b^{e,n}$ at node *n* [10], [17]. The binary decision variable $u^{e,n}$ is used to model the impact of PSPS decisions on line power flows. Constraint (A10) enforces thermal limits of the line, and sets flow to zero when a line is de-energized.

$$P^{e,n} \leq -b^{e,n}(\theta^{i,n} - \theta^{j,n} + \theta^{max}(1 - u^{e,n})) \quad \forall i \in \beta, \forall j \in \beta, e \in E(i,j) \quad \text{(A8)}$$

$$P^{e,n} \geq -b^{e,n}(\theta^{i,n} - \theta^{j,n} + \theta^{min}(1 - u^{e,n})) \quad \forall i \in \beta, \forall j \in \beta, e \in E(i,j) \quad \text{(A9)}$$

$$-T^e u^{e,n} \leq P^{e,n} \leq T^e u^{e,n} \quad \text{(A10)}$$

Constraint (A11) ensures power balance at each bus, requiring total injections to equal total withdrawals. The variable $l^{d,n}$ captures the impact of PSPS decisions on load in the next stage.

$$\sum_{g \in B_i^G} P^{g,n} = \sum_{e \in B_i^e} P^{e,n} + \sum_{d \in B_i^d} l^{d,n} D^{d,n} \quad \forall i \in \beta, e \in E(i,j) \quad \text{(A11)}$$

Equation (A12) models generator *g*'s active power at node *n* by adjusting its previous stage output $P^{g,a(n)}$ in decision tree. This constraint captures the generation adjustment for each generator by applying a differential term $\Delta P^{g,n}$ to its prior stage generation level $P^{g,a(n)}$, where $a(n)$ represents the parent node of *n*.

$$P^{g,n} = P^{g,a(n)} + \Delta P^{g,n} \quad \forall g \in G \quad \text{(A12)}$$

$P^{g,n}$ in (A12) are regarded as state variables for power generation, playing a key role in the multistage optimization process. Equation (A13) explicitly defines the generator ramping rate constraints, ensuring that the change in output between consecutive stages does not exceed the allowable ramp-up or ramp-down rates. These ramping rate limits are essential for maintaining operational feasibility during transitions across stages.

$$\Delta p_{min}^g \leq \Delta P^{g,n} \leq \Delta p_{max}^g \quad \forall g \in G \quad \text{(A13)}$$

Constraint (A14) establishes the lower and upper limits for real power generation by each generator at every node. The binary decision variable $u^{g,n}$ represents the effect of PSPS actions, which may require shutting down some generators.

$$u^{g,n} P_{min}^g \leq P^{g,n} \leq u^{g,n} P_{max}^g \quad \forall g \in G \quad \text{(A14)}$$